\documentclass[twocolumn]{aastex62}

\newcommand{\gadget}{\textsc{Gadget}}
\newcommand{\sphgal}{\textsc{SPHGal}}
\newcommand{\subfind}{\textsc{Subfind}}
\newcommand{\treecol}{\textsc{TreeCol}}

\graphicspath{{./}{figures/}}

\received{24 May 2019}
\revised{13 June 2019}
\accepted{14 June 2019}
\submitjournal{ApJL}

\shorttitle{}
\shortauthors{Lah\'{e}n et al.}

\begin{document}

\title{The formation of low metallicity globular clusters in dwarf galaxy mergers}

\correspondingauthor{Natalia Lah\'{e}n}
\email{natalia.lahen@helsinki.fi}

\author[0000-0003-2166-1935]{Natalia Lah\'{e}n}
\affil{Department of Physics, University of Helsinki, Gustaf H\"allstr\"omin katu 2, FI-00014 Helsinki, Finland}

\author{Thorsten Naab}
\affiliation{Max Planck Institute for Astrophysics, Karl-Schwarzschild-Str. 1, D-85740, Garching, Germany}

\author{Peter H. Johansson}
\affiliation{Department of Physics, University of Helsinki, Gustaf H\"allstr\"omin katu 2, FI-00014 Helsinki, Finland}

\author{Bruce Elmegreen}
\affiliation{IBM T. J. Watson Research Center, 1101 Kitchawan Road, Yorktown Heights, NY 10598, USA}

\author{Chia-Yu Hu}
\affiliation{Center for Computational Astrophysics, Flatiron Institute, 162 5th Ave, New York, NY 10010, USA}

\author{Stefanie Walch}
\affiliation{I. Physikalisches Institut, Universit\"at zu K\"oln, Z\"ulpicher Strasse 77, D-50937 K\"oln, Germany}
\affiliation{Cologne Center for Data and Simulation Science, University of Cologne, www.cds.uni-koeln.de}

\begin{abstract}
We present a hydro-dynamical simulation at sub-parsec and few solar mass resolution of a merger between two gas-rich dwarf galaxies. Our simulation includes a detailed model for the multi-phase interstellar medium (ISM) and is able to follow the entire formation history of spatially resolved star clusters, including feedback from individual massive stars. Shortly after the merger we find a population of $\sim 900$ stellar clusters with masses above $10^{2.5}\; \rm{M_\odot}$ and a cluster mass function (CMF), which is well fitted with a power-law with a slope of $\alpha=-1.70\pm0.08$. We describe here in detail the formation of the three most massive clusters (M$_{*} \gtrsim 10^5$ M$_\odot$), which populate the high-mass end of the CMF.
The simulated clusters form rapidly on a timescale of $6$--$8$ Myr in converging flows of dense gas. The embedded merger phase has extremely high star formation rate surface densities of $\Sigma_\mathrm{SFR}>10\; \mathrm{M}_\odot\; \mathrm{yr}^{-1}\; \mathrm{kpc}^{-2}$ and thermal gas pressures in excess of $P_{\rm th}\sim10^7 \; \mathrm{k}_{\rm B}\;(\rm K\;\mathrm{cm}^{-3})^{-1}$. The formation process is terminated by rapid gas expulsion driven by the first generation of supernovae, after which the cluster centers relax and both their structure and kinematics become indistinguishable from observed local globular clusters. 
The simulation presented here provides a general model for the formation of metal-poor globular clusters in chemically unevolved starbursting environments of low-mass dwarf galaxies, which are common at high redshifts.

\end{abstract}


\section{Introduction} \label{sec:intro}

Globular clusters (GCs) are the densest gravitationally bound stellar systems in the Universe, and 
they are found in all types of galaxies, even low-mass dwarf galaxies \citep{1991ARA&A..29..543H}. GCs have
old ages ($\gtrsim 100$ Myr), with some of the oldest clusters being nearly as old as the Universe \citep{2009ApJ...694.1498M}, and they have typical masses of several times $10^5$ M$_\odot$, with typical effective radii of several parsecs. The elemental abundances of GCs range from $0.3\%$ solar to about solar metallicity  and they show diverse chemical compositions indicative of multiple stellar populations (e.g. \citealt{2018ARA&A..56...83B}).  

It is very challenging to resolve the formation sites of GCs at high redshifts with present-day telescopes, making their star formation environments largely unknown (e.g. \citealt{2017MNRAS.467.4304V}). However, the present-day structure of GCs suggests that extreme gas pressures, about $\sim 2-3$ orders of magnitude higher than in the present-day Milky Way, were required to make them at their high stellar densities and gravitational binding energy densities \citep{1997ApJ...480..235E,2012MNRAS.426.3008K}. 

Contrary to the high pressure requirement, present-day dwarf galaxies typically have the lowest interstellar pressures (e.g. \citealt{2000ApJ...540..814E}).  This contradiction has led to the idea that at least some metal-poor GCs formed during mergers of dwarf galaxies, where fast shock fronts compressed the gas and caused large-scale gravitational collapse into dense clouds \citep{1991ApJ...379..157B}. 

Numerically the formation of massive stellar clusters have been studied using both isolated simulations of collapsing giant molecular clouds (e.g. \citealt{2011MNRAS.414..321D,2018NatAs...2..725H}), in mergers of both spiral (e.g. \citealt{2002MNRAS.335.1176B,2008MNRAS.389L...8B}) and dwarf galaxies (e.g. \citealt{2010ASPC..423..185S,2013MNRAS.430.1901H}) and recently also in cosmological simulations (e.g. \citealt{2016ApJ...823...52K,2018MNRAS.474.4232K}).

The pioneering simulations of \citet{2002MNRAS.335.1176B} and \citet{2008MNRAS.389L...8B} resolved the 
forming stellar clusters with particle masses of $\sim 7000 \; \mathrm{M}_{\odot}$ at a typical spatial resolution of a few hundred parsecs.
More recently, improved numerical merger simulations by \citet{2013MNRAS.430.1901H} and \citet{2015MNRAS.446.2038R} resolved masses down to $\sim 100\; \mathrm{M}_{\odot}$ at $\sim 1$ parsec scales, whereas current galaxy-scale simulations of isolated dwarf galaxies are even able to resolve the ISM down to sub-parsec scales and follow the feedback from individual stars (e.g. \citealt{2017MNRAS.471.2151H,2018MNRAS.480.1666S,2019MNRAS.482.1304E}). 

The formation of a GCs is fundamentally a two-step process \citep{2014CQGra..31x4006K}. Firstly, enough mass must be accumulated to form a massive stellar cluster and secondly this cluster must survive for at least $\sim 100$ Myr and up to a Hubble time in order to be observable in the present-day Universe. In this Letter we concentrate on the formation phase of massive stellar clusters, and present a hydro-dynamical simulation of a dwarf galaxy merger resolved at stellar-mass and sub-parsec spatial resolution.

\section{Simulation setup}

\subsection{Simulation code}

The simulations were run using a modern smoothed particle hydrodynamics (SPH) solver \sphgal\ \citep{2014MNRAS.443.1173H,2016MNRAS.458.3528H} based on the \gadget\ -code  \citep{2005MNRAS.364.1105S}.
The code employs a pressure-energy formulation of the hydrodynamic equations \citep{2013MNRAS.428.2840H} using the Wendland C$^4$ smoothing kernel \citep{2012MNRAS.425.1068D} smoothed over 100 neighbors, and also includes well-tested implementations of artificial viscosity and thermal conduction in converging gas flows \citep{2008JCoPh.22710040P}. 

In the temperature range $10 \ \mathrm{K} <T<3\times 10^4 \ \rm K$ gas is cooled using the model described in \citet{2016MNRAS.458.3528H}, which is based on a non-equilibrium chemical network \citep{2007ApJS..169..239G} that follows six chemical species (H$_{2}$, H$^{+}$, H, CO, C$^{+}$, O), in addition to the free electron density. For the dust-to-gas mass ratio we assume a constant value of $0.1\%$. At temperatures above $T>3\times 10^4 \ \rm K$, we adopt an equilibrium cooling model using a metallicity-dependent cooling function \citep{2009MNRAS.393...99W}. Metallicity in the ISM is followed by tracing the mass of $12$ elements (H, He, N, C, O, Si, Mg, Fe, S, Ca, Ne, Zn) in both gas and star particles \citep{2013MNRAS.434.3142A}, which evolve through stellar feedback.  In this temperature regime the ISM is assumed to be optically thin and in ionisation equilibrium with a cosmic UV background \citep{2001cghr.confE..64H}.

We follow the mass, metal \citep{2004ApJ...608..405C} and radiation \citep{2016MNRAS.458.3528H, 2017MNRAS.471.2151H} output into the ISM along the lifetime of each massive star. 
The stellar radiation includes the far-ultraviolet (FUV) radiation with photon energy below $13.6$ eV and the hydrogen-ionising radiation from massive stars. Shielding against the FUV radiation is accounted for by using the \treecol\ algorithm \citep{2012MNRAS.420..745C}.

The star formation model samples stellar masses from a Kroupa \citep{2001MNRAS.322..231K} initial mass function  and statistically realises each individual star more massive than $4$ M$_\odot$ as a stellar particle.
If the sampled mass exceeds the available mass in the gas particle, the mass deficit is stochastically taken from nearby gas particles in order to satisfy mass conservation. Stars with masses below $4$ M$_\odot$ are grouped into stellar population particles.

\begin{figure}[h!]
\includegraphics[width=\columnwidth]{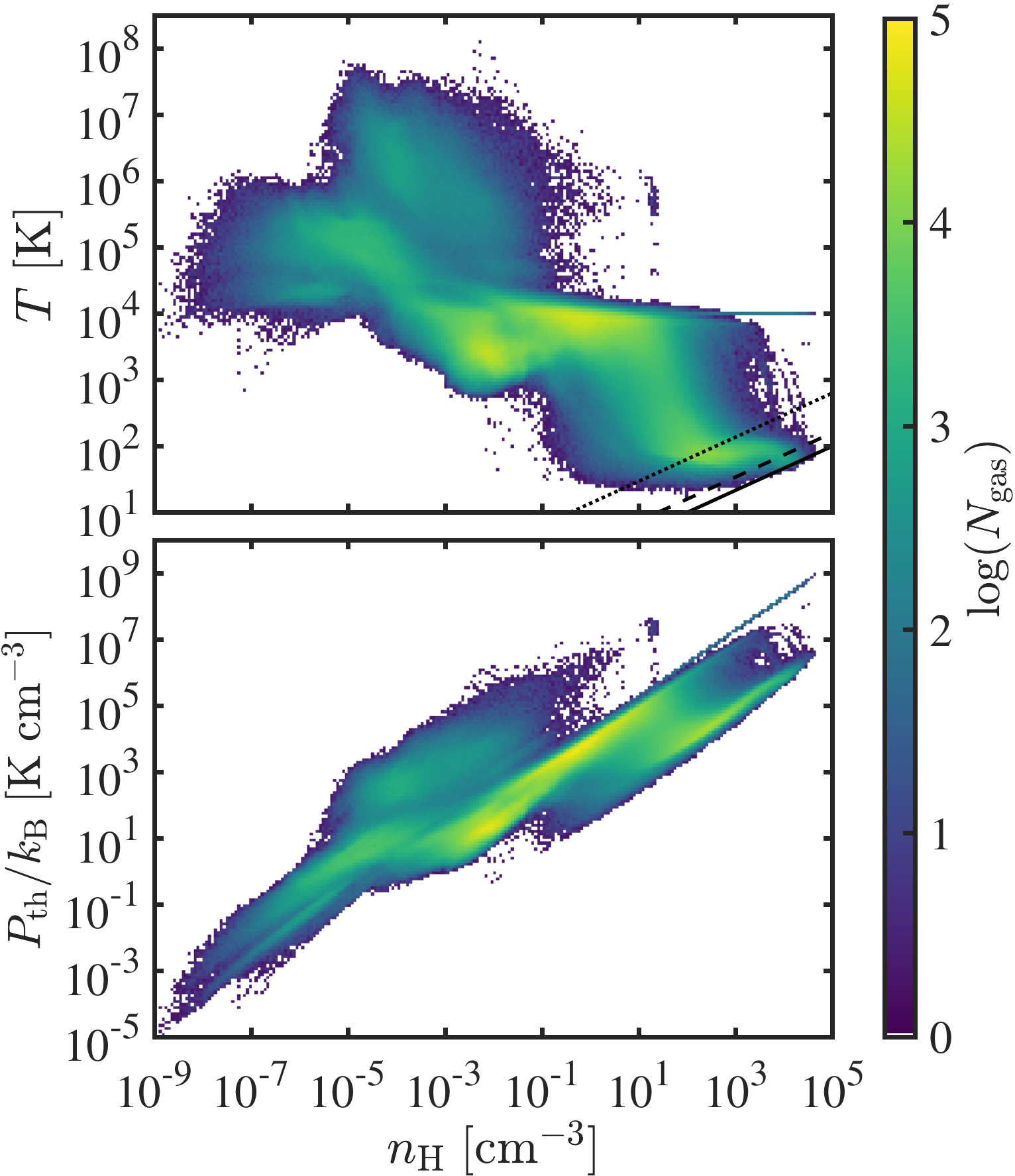}
 \caption{Gas phase diagrams showing temperature (top) and thermal pressure (bottom) as a function of hydrogen number density at the time of peak SFR at $t_{0}- 5$ Myr. The SF thresholds are shown with diagonal dotted and solid lines which correspond to the onset of SF at $M_{\rm J}=8$ kernel masses at an efficiency of $2\%$ and enforced SF at $0.5$ kernel masses, respectively. The dashed line indicates where the SPH kernel mass ($\sim$ 400 M$_\odot$) equals one Jeans mass.
 \label{phasediagram} }
\end{figure}

Star formation (SF) is taking place at a $2\%$ efficiency on a dynamical time-scale for gas particles which are present in a converging flow and have a Jeans mass
\begin{equation}
 M_{J} = \frac{\pi^{5/2}c_{s}^3}{6G^{3/2}\rho^{1/2}}
\end{equation}
below eight SPH kernel masses, where $G$ is the gravitational constant, $\rho$ is gas density, and $c_s$ is the local sound speed of the gas. In addition, we enforce instantaneous star formation when the Jeans mass falls below half a kernel mass, i.e. $\sim 200$ M$_\odot$ (see Fig. \ref{phasediagram}).

At the end of their lifetimes, stars more massive than $8$ M$_\odot$ undergo type II supernova (SNII) explosions with mass-dependent delay times \citep{2017MNRAS.471.2151H}, with the resulting mass and metal yields distributed within the weighted SPH kernel. 
At this low metallicity and dust content, feedback from supernovae and ionising radiation will dominate and the additional contribution from radiation pressure on dust (e.g. \citealt{2018NatAs...2..725H}) and from line driven winds (e.g. \citealt{2008A&ARv..16..209P}) are expected to be weak and are therefore not included in our present model.

\begin{figure*}
\includegraphics[width=\textwidth]{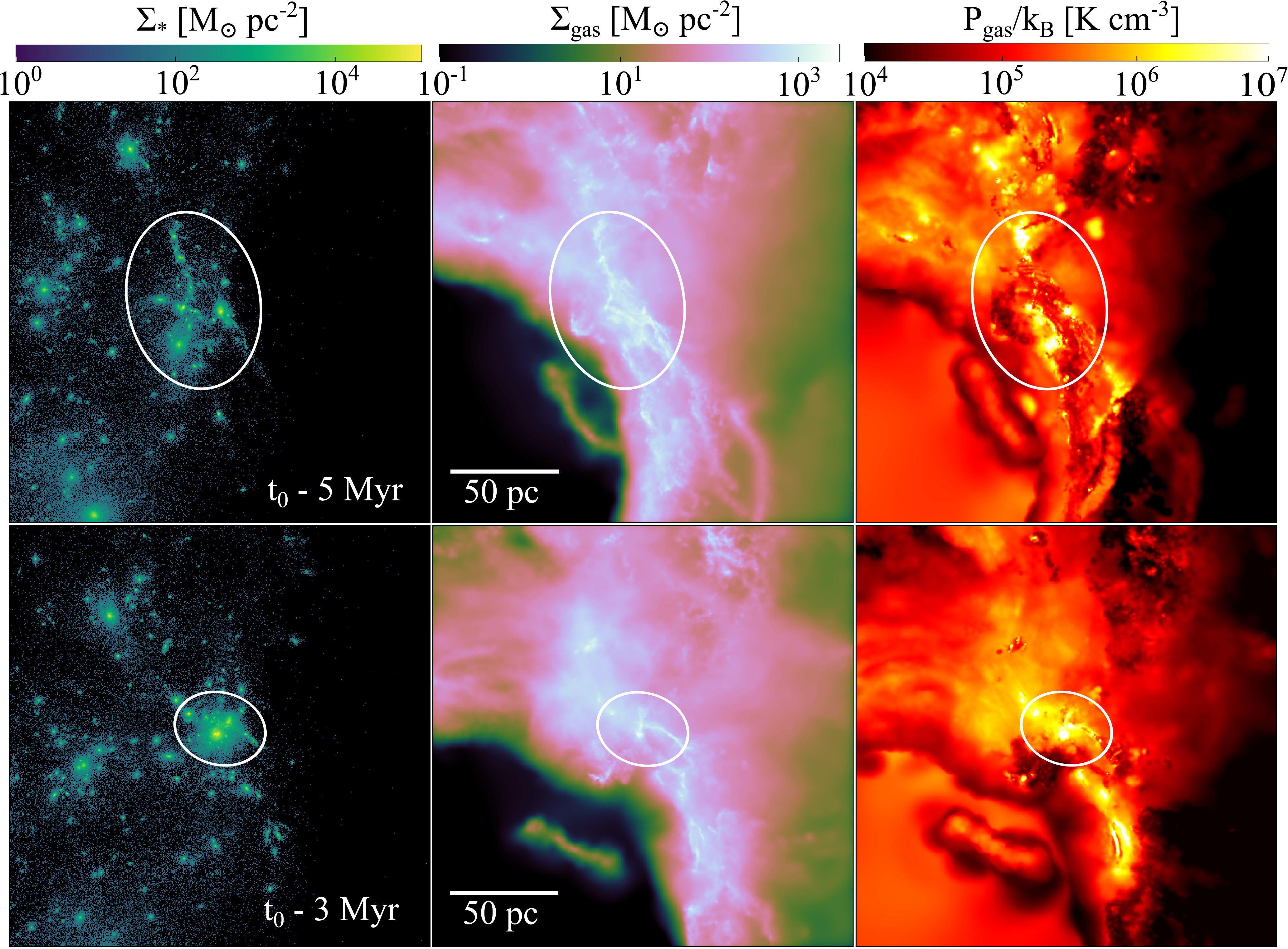}
 \caption{The surface density of stars (left) and gas (middle), and thermal gas pressure (right) in a $200$ pc cube centered at the formation region of the most massive cluster (white ellipses). The top and bottom rows show the region $5$ Myr and $3$ Myr before the final cluster assembly at $t_{0}$, respectively.
 Tens of stellar clusters are forming within the filamentary gas structures, with additional compression provided by the super-shell generated from the first forming massive stellar cluster. \label{fig1}}
\end{figure*}

\subsection{Initial conditions}

The initial conditions consist of two identical dwarf galaxies
setup at z=0 with virial radii of $44$ kpc  including identical dark matter halos, stellar disks and extended gas disks. The dark matter halo is set up with a Hernquist profile  
with virial mass of $M_{\rm vir} = 2\times 10^{10}$ M$_\odot$, an NFW-equivalent \citep{2005MNRAS.361..776S} concentration parameter of $c=10$ and a Bullock spin-parameter of $\lambda=0.03$. 

The rotationally supported stellar disk with mass $M_*=2\times 10^{7}\; \mathrm{M}_{\odot}$ has an exponential surface density distribution with a scale radius of $r_{\rm D,star}=0.73 \ \rm kpc$ and a constant scale height of $h_{\rm star}=0.35 \ \rm kpc$. The exponential gas disk with a mass of $M_{\rm gas}=4\times 10^{7}\; \mathrm{M}_{\odot}$ is set up with a scale radius of $r_{\rm D,gas}=1.46 \ \rm kpc$ and a gas scale height initially set by the condition of hydrostatic equilibrium.
The initial metallicity of the gas disk is set to a constant value of $0.1$ Z$_\odot$.

The galaxies are oriented off the orbital plane with the inclination and argument of pericenter chosen as $\{i_{1}, i_{2}\}=\{60^{\circ}, 60^{\circ}\}$ and $\{\omega_{1}, \omega_{2}\}=\{30^{\circ}, 60^{\circ}\}$.
The parabolic orbit of the dwarfs, with an initial distance of $5$ kpc and a pericentric separation of $1.46$ kpc, results in a relatively rapid coalescence time for the two gas disks. 
The first passage occurs $\sim 50$ Myr from the start of the simulation, followed by the second encounter $\sim 100$ Myr later, with the final coalescence of the gas disks occurring $\sim 20$ Myr after the second passage. The simulation was run for more than $100$ Myr past the coalescence for a total simulation time of $\sim300$ Myr. 

The gravitational forces are softened with a spline kernel with a softening length of $\epsilon=0.1 \ \rm pc$ for stars and gas, and $\epsilon=62 \ \rm pc$ for dark matter, respectively. The time evolution is resolved down to $\sim 10$ yr. The mass resolution for the stellar and gaseous particles is typically $4$ M$_\odot$ per particle. The dark matter component has a mass resolution of $10^4$ M$_\odot$ per particle.  

Our simulation is limited by our gravitational softening length of $0.1$ pc and mass resolution of $\gtrsim 4$ M$_\odot$. In this study we thus concentrate on the formation and early evolution of GCs. However, we stress that unsoftened simulations with direct N-body codes have shown that massive clusters similar to ours evolve into present-day GCs after $10$ Gyrs of evolution, even in the tidal field of the Milky Way \citep{2016MNRAS.458.1450W}. 

\begin{figure}
\includegraphics[width=\columnwidth]{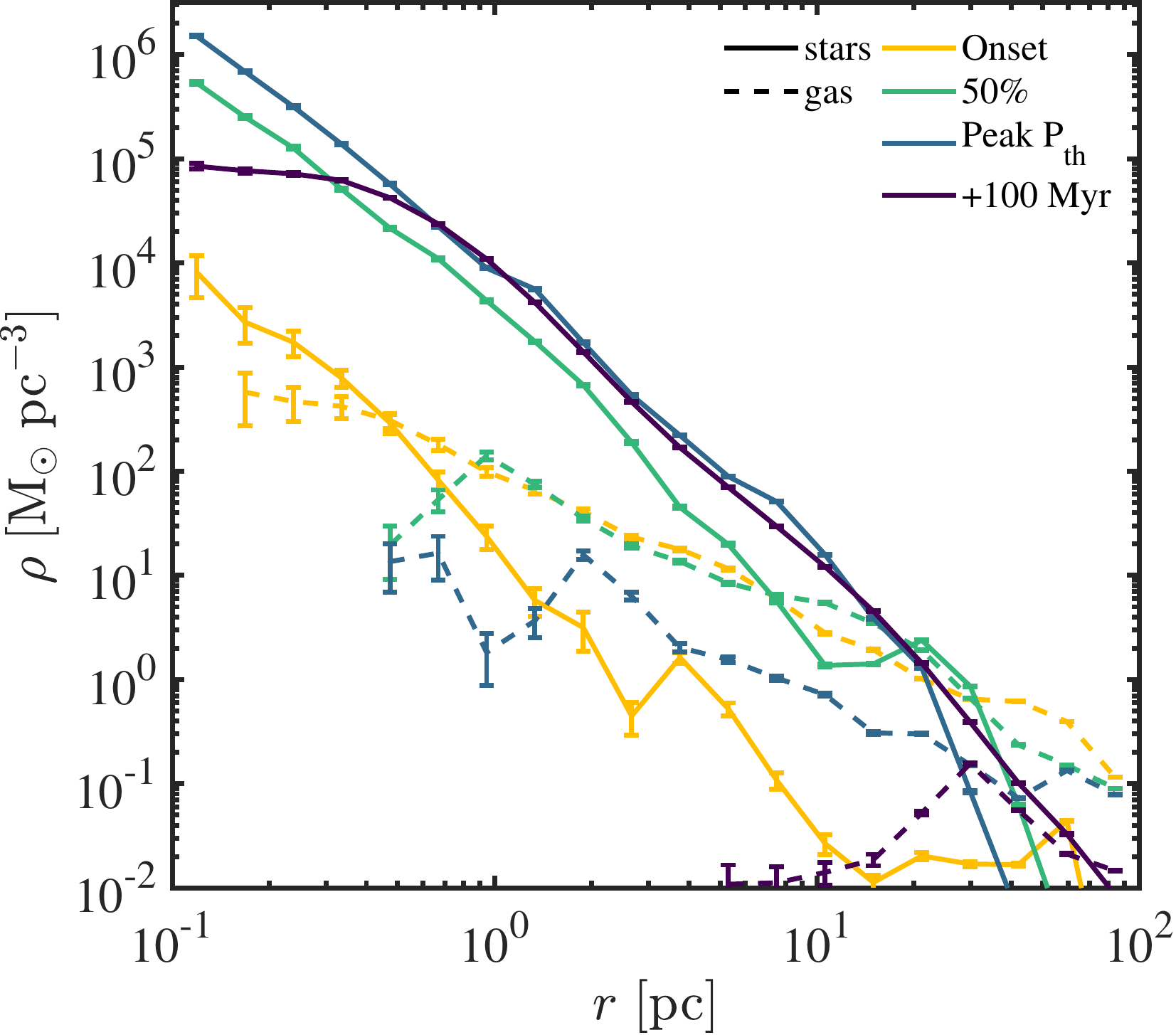}
 \caption{The radial mass density profiles of stars (solid) and cold gas (T$<10^{2.5}$ K, dashed) within the most massive stellar cluster, shown at the onset of cluster formation at $t_0-10$ Myr (yellow), the time when $50\%$ of the cluster mass has formed at $t_0-5$ Myr (green), the time when the thermal pressure peaks in the region at $t_0-3$ Myr (blue), and $100$ Myr after the cluster formation (purple). The plotted errors correspond to one bootstrapped standard deviation.
 \label{densities} }
\end{figure}

\section{Results}

\subsection{The formation of stellar clusters}
When evolved in isolation, the model dwarf galaxies have a low SFR of $10^{-3}\; \mathrm{M}_\odot\; \mathrm{yr}^{-1}$, in agreement with the long gas depletion times observed locally in these types of systems \citep{2017MNRAS.471.2151H}. However, during the galaxy merger the global SFR increases by two orders of magnitude to SFR $=0.3$ M$_\odot$ yr$^{-1}$, with a concurrent increase in the fraction of stars forming in bound star clusters (see e.g. \citealt{2012MNRAS.426.3008K}). 

In Fig. \ref{phasediagram} we show the phase diagrams for all gas in the merging dwarf system at the time of peak SFR. The instantaneous SF threshold (solid line) corresponds to densities of $n_{\rm H}=10^{3.5} - 10^{4.5} \ \rm cm^{-3}$, at temperatures in the range of $T=10 - 100\ \rm K$, with the most extreme values only reached during the merger.

We use the \subfind\ algorithm \citep{2001MNRAS.328..726S} to identify stellar clusters in the simulation. 
In this study we define clusters to be objects with more than $50$ bound stellar particles, which results in a minimum cluster mass of $M_*\sim 200$ M$_\odot$.

After the coalescence of the galactic disks the merger remnant harbors a considerable population of more than 1000 young stellar clusters. The distribution of the $\sim 900$ clusters with masses above a typical observed cut-off of $M_*>10^{2.5}\; \mathrm{M}_\odot$ follows a mass function of the form  $dN\propto M^{\alpha} dM$ with a best-fit power-law index of $\alpha =-1.70 \pm 0.08$, as also seen in observed dwarf galaxies (e.g. \citealt{2010ApJ...711.1263C}). At the peak epoch of star formation we identify three compact regions with extremely high SFR surface densities exceeding $\Sigma_{\rm SFR}>10$ M$_\odot$ yr$^{-1}$ kpc$^{-2}$, comparable to values in massive high-redshift starburst galaxies \citep{2013ApJ...768...74T}. Each of these regions produces a massive dense star cluster on a timescale of only a few Myr, after which the clusters remain on orbits with 
apocenters of $\sim 500$ pc from the center of mass of the merger remnant, similarly to the globular clusters in local dwarf galaxies \citep{2003MNRAS.338...85M}.

We show in Fig. \ref{fig1} the environment of the most massive cluster, where the average gas surface density exceeds $\Sigma_{\rm gas} \sim 10^2$ M$_\odot$ pc$^{-2}$ with thermal pressures of $P_{\rm th} \sim 10^5$--$10^7$ k$_\mathrm{B} \;(\rm K\;\mathrm{cm}^{-3})^{-1}$. 
The top panels depict the region when the cluster has assembled half of its final stellar mass, $5$ Myr before the formation is completed at time t$_0$. The bottom panels show the time when the maximum thermal pressure is reached, approximately $2$ Myr later. At these times the entire formation region is still embedded in a dense gas filament. Over a total time span of $\sim 6$--$8$ Myr, nearby newly formed stellar clusters with a wide range of masses along the CMF merge together while embedded in a converging gas flow that funnels gas at a typical infall rate of $0.2\; \mathrm{M}_\odot \; \mathrm{yr}^{-1}$ into a region $50$ pc in radius.

The slow rotation of the galaxies allows for such a prolonged collapse phase in the absence of destructive shear forces. Additional compression is provided by a super-shell generated by the first massive cluster that formed. The central density of gas within the cluster formation region reaches $10^3$--$10^4$ M$_\odot$ pc$^{-3}$ ( $n_{\rm H}>10^4$ cm$^{-3}$) where the gas is efficiently transforming into stars on the local dynamical timescale of $\tau_{\mathrm{dyn}} \sim 10^6$ yr. The mean SFR in this region is $0.1$ M$_\odot$ yr$^{-1}$ at a mean efficiency of $\epsilon_{\rm{SF}}\sim \dot{M}_{\rm *}/\dot{M}_{\rm infall}\sim 50\%$ with respect to the infalling gas, with peak values of up to $\epsilon_{\rm{SF}}\sim90\%$.

\begin{figure*}
\includegraphics[width=\textwidth]{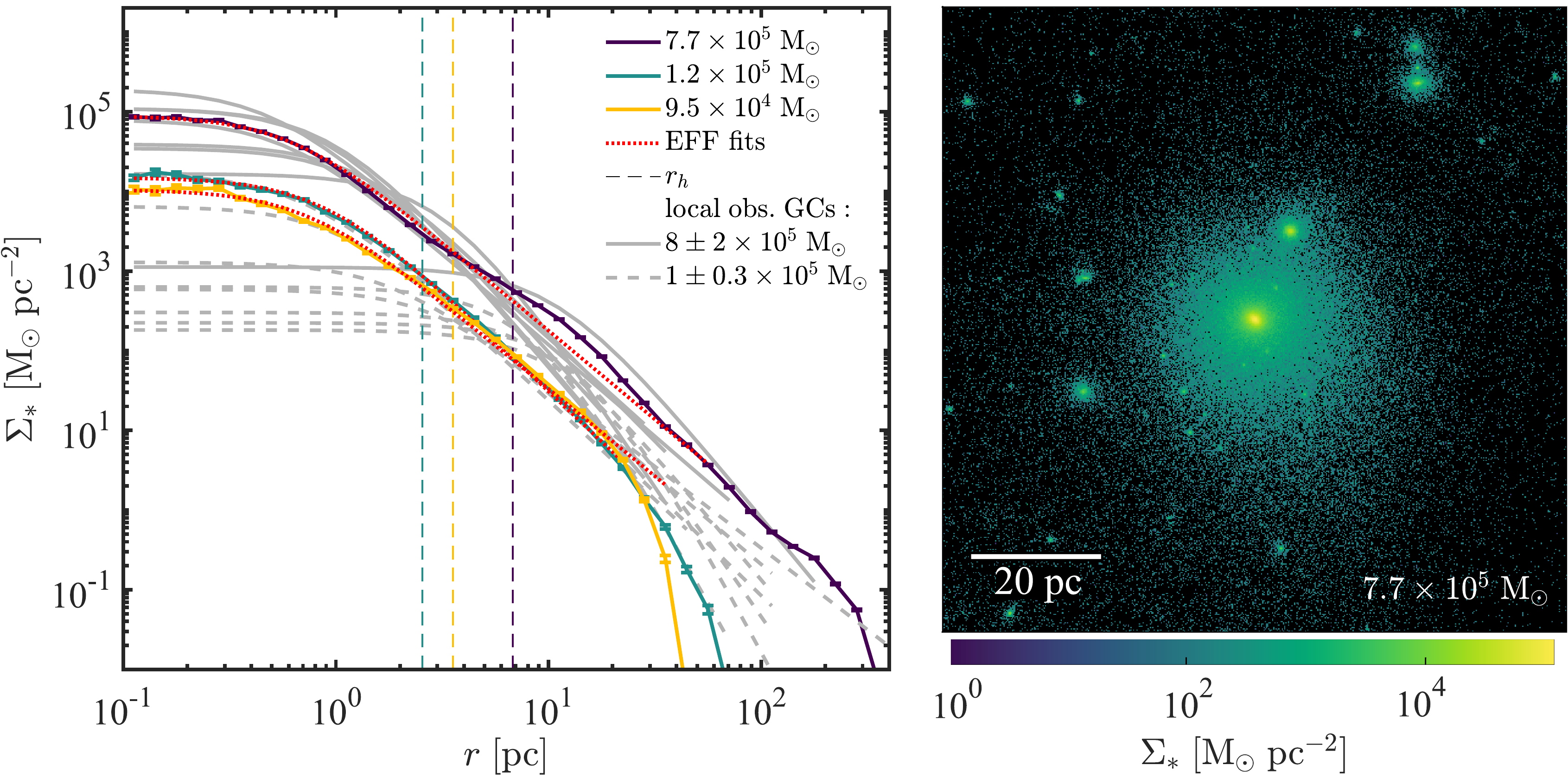}
 \caption{Left panel:  Radial stellar surface density profiles of gravitationally bound stars in the three most massive star clusters $100$ Myr after their formation. The half-mass radii ($r_h$) and best-fit EFF profiles  (Eq. \ref{eq:EFF}) are over-plotted. The simulated surface density profiles agree well with the best-fit EFF profiles of $8$ GCs in the Milky Way (solid grey) with stellar masses of $8\pm 2 \times 10^5 \; \mathrm{M}_\odot$ and with $5$, $1$, and $2$ GCs in the LMC, SMC, and Fornax (dashed grey), respectively, with masses of $1 \pm 0.3 \times$ $10^5\; \mathrm{M}_\odot$ \citep{2005ApJS..161..304M}. Right panel: The stellar surface density map of $\sim 2\times 10^5$ stellar particles gravitationally bound to the most massive globular cluster, and its immediate stellar environment $100$ Myr after its formation. \label{surface_density} }
\end{figure*}

\subsection{The mass assembly of stellar clusters}\label{sec:build_up}

Fig. \ref{densities} shows the rapid build-up of stellar density in the most massive cluster, where the newly formed stars concentrate at stellar densities of $\rho_{*} > 10^6$ M$_\odot$ pc$^{-3}$. This maximum stellar density exceeds the density of the gas by more than two orders of magnitude and has a steeper radial density profile, in good agreement with what is observed in young massive clusters \citep{2016MNRAS.457.4536W}. 

The central 10 half-mass radii (see Section \ref{sec:cluster_props}) around the center of mass of each of the three most massive clusters retains a fairly constant gas infall rate of $0.05$--$0.2\; \mathrm{M}_\odot \; \mathrm{yr}^{-1}$ during their assembly, while on average $25\%$--$50\%$ of the infalling gas mass is converted into stars which mostly end up in the central cluster in each region. 

Following the increasing SFR, the SNII rate increases as the cluster mass builds up. Typically, not more than $10\%$ of the supernova energy couples to the surrounding dense ISM even if the medium is ionised \citep{2015MNRAS.451.2757W}. However, at high enough supernova rates ($\sim 500$ Myr$^{-1}$), even compensating for the short cooling times of $\tau_{\rm cool} \sim 2 \times 10^3$ yr in the dense gas, the supernovae end up depositing as much as the gas binding energy of $4 \times 10^{52}$ erg. As a result, all of the remaining gas is blown out of the cluster about $7$ Myr into its formation, soon after the epoch shown in the bottom panel of Fig. \ref{fig1}. 

At this time the most massive cluster has reached a mass of  $\sim 8\times10^5$ M$_\odot$ and has consumed $\sim 30\%$ of the available gas within the nearby $50$ pc region; the remaining gas has either been expelled ($63\%$) or converted into stars ($7\%$) mostly bound in smaller mass clusters ($5\%$). By the time the gas is expelled, about $80\%$ of the cluster stars have formed in-situ within $30$ pc, with the remaining $20\%$ of the stellar mass coming from smaller clusters and unbound stars that accreted from the surrounding $100$ pc neighborhood.

\subsection{Comparison of cluster properties with observations}\label{sec:cluster_props}

Once star formation is terminated, the cluster evolution is governed by its tidal environment and by collisional processes such as two-body relaxation. Even in a softened system, the number of particles is the main driver of the two-body relaxation timescale \citep{2008gady.book.....B}. At our gravitational softening of $\epsilon=0.1 \ \rm pc$ we find that a constant density stellar core forms roughly on the local relaxation timescale of $t_{\rm relax}\sim 40$ Myr, which is within a factor of a few of the true relaxation time given our stellar-mass particle resolution.

In Fig. \ref{surface_density} we show the radial and 2D stellar surface density profiles of the most massive cluster $\sim 100$ Myr after assembly, together with the radial surface densities of the two next-most massive clusters, which have similar formation histories. Evolving the simulated clusters several Gyrs would further lower the central stellar density \citep{2016MNRAS.458.1450W}, bringing especially the lower-mass clusters closer to their observed counterparts.

A common model for fitting the surface brightness and mass profiles of especially young and intermediate-age stellar clusters, is the Elson-Fall-Freeman power-law (EFF, \citealt{1987ApJ...323...54E,2005ApJS..161..304M}) which is defined as
\begin{equation}\label{eq:EFF}
    \Sigma(r)=\Sigma_0 \left(1+ \frac{r^2}{a^2} \right)^{-\gamma/2}
\end{equation}
and parametrised by the central surface density $\Sigma_0$, the scale radius $a$ and the power-law index $\gamma$. 
All three simulated clusters in Fig. \ref{surface_density} resemble the EFF profiles observed for similar mass clusters with ages in excess of $100$ Myr in the Local Group \citep{2005ApJS..161..304M}, and they are well fitted by similar power-law exponents of $\gamma\sim-2.2$ to $\gamma\sim-2.5$ above the limit of $\Sigma_{*} \sim 10$ M$_\odot$ pc$^{-2}$
set by the ambient stellar background. 

\begin{figure}
\includegraphics[width=\columnwidth]{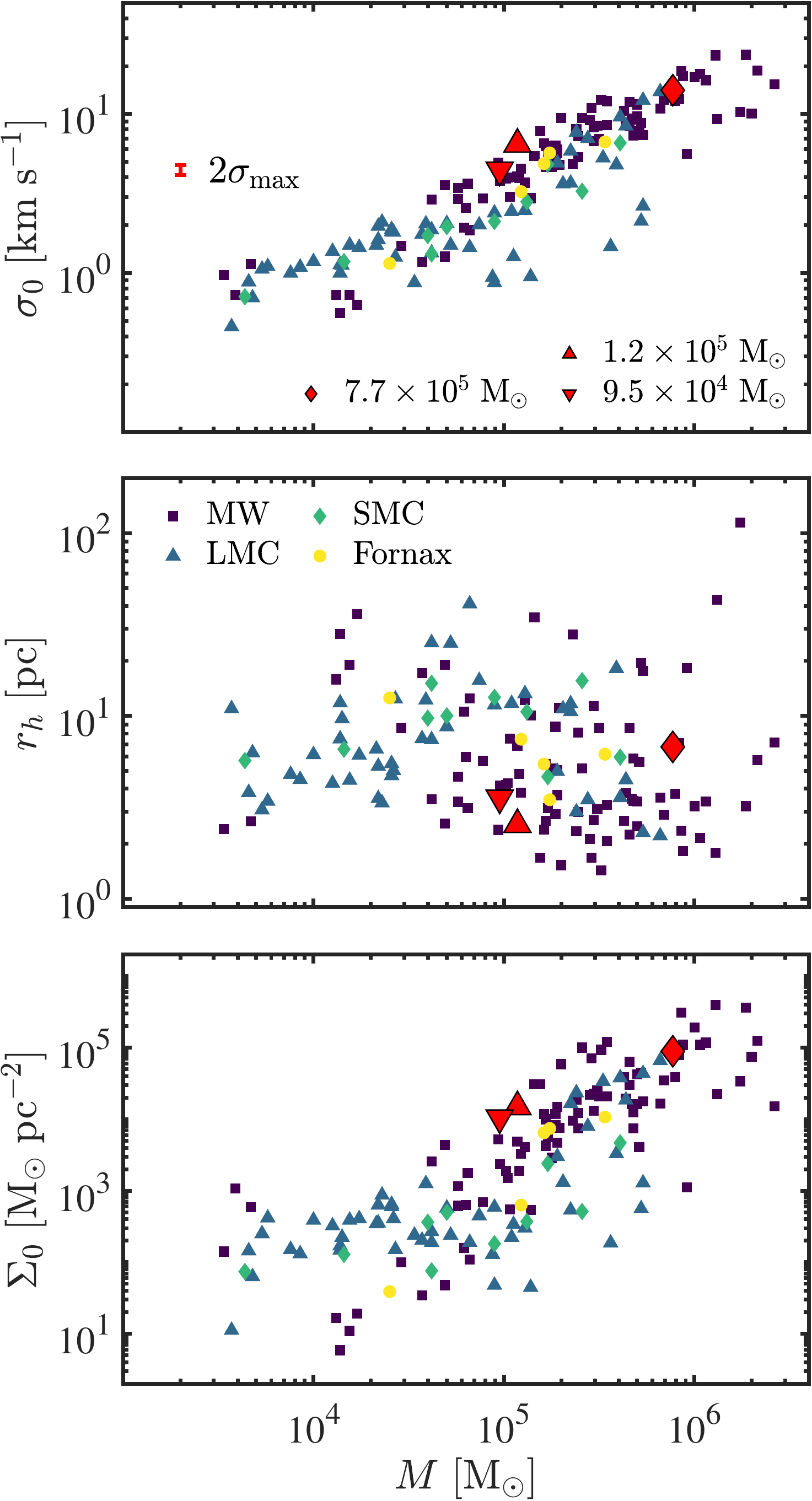}
 \caption{ Structural properties of the three simulated GCs (red symbols) compared to 153 observed massive stellar clusters in the Local Group \citep{2005ApJS..161..304M}. From top to bottom we show the central velocity dispersion, simulated half-mass (observed half-light) radius, and central stellar surface density. For observed quantities we show values from the best-fit EFF profiles. In the top panel we also show the maximum error from a bootstrapping analysis. 
 \label{scaling_relations}}
\end{figure}

In Fig. \ref{scaling_relations} we show how the velocity dispersions, half-mass radii, and central surface densities of the three model GCs are indistinguishable from massive stellar clusters in the dwarf galaxies Fornax and the Small and Large Magellanic Clouds (SMC and LMC), and the Milky Way \citep{2005ApJS..161..304M}. Out of the cluster sample shown in Fig. \ref{scaling_relations}, only the SMC (1 cluster) and LMC (16 clusters) harbor stellar clusters younger than 100 Myr, respectively. However, the properties of these younger clusters are inseparable from the other older clusters in the sample shown in Fig. \ref{scaling_relations}. 
The average stellar metallicities of the simulated clusters are close to the initial value of $0.1\; Z_{ \odot}$, indicating that the clusters have not been significantly enriched by star formation during their short formation times. The scatter in stellar ages for the simulated GCs is $4$ Myr or less, also in good agreement with the observations.

\section{Conclusions}

We have presented results from a detailed numerical simulation in which three dense massive stellar clusters form on timescales of several Myr in a dwarf galaxy merger with low metallicity.  After the gas is expelled by supernovae, the clusters relax and appear similar to present-day GCs.

We propose that such a formation scenario results in GCs in chemically unevolved, low-mass galaxies (M$_{*}<10^9$ M$_\odot$) at high redshift ($z > 3.5$) (e.g. \citealt{2012ApJ...757....9E}). The peak SFR of $0.3$ M$_\odot$ yr$^{-1}$ corresponds to a relatively low H$\alpha$ luminosity \citep{2010AJ....139..447H} of L$_{\rm{H}_\alpha}=4.3\times 10^{40}$ erg s$^{-1}$. The most massive clusters can either survive to the present-day in these dwarfs, where tidal forces are weak or, alternatively, they can be accreted onto more massive systems such as the Milky Way \citep{2010ApJ...718.1266M}. Finally, this formation scenario is also consistent with observations of young massive clusters in the local Universe, which are thought to be the counterparts of old globular clusters \citep{2010ARA&A..48..431P}.

\small
\begin{acknowledgements}
The computations were carried out at CSC -- IT Center for Science Ltd. in Finland and at Max-Planck Institute for Astrophysics in Germany.
N.L. acknowledges the financial support by the Jenny and Antti Wihuri Foundation. N.L. and P.H.J. acknowledges support by the Academy of Finland grant 274931 and the European Research Council via ERC Consolidator Grant KETJU (no. 818930). C.-Y.H. acknowledges The Center for Computational Astrophysics, supported by the Simons Foundation. S.W. acknowledges support by the European Research Council via ERC Starting Grant RADFEEDBACK (no. 679852) and by the German Science Foundation via CRC956, Project C5.

\end{acknowledgements}

\end{document}